\documentclass{iaus}
\usepackage{graphics}

  \checkfont{eurm10}
  \iffontfound
    \IfFileExists{upmath.sty}
      {\typeout{^^JFound AMS Euler Roman fonts on the system,
                   using the 'upmath' package.^^J}%
       \usepackage{upmath}}
      {\typeout{^^JFound AMS Euler Roman fonts on the system, but you
                   dont seem to have the}%
       \typeout{'upmath' package installed. iaus.cls can take advantage
                 of these fonts,^^Jif you use 'upmath' package.^^J}%
      }
  \else
  \fi

  \checkfont{msam10}
  \iffontfound
    \IfFileExists{amssymb.sty}
      {\typeout{^^JFound AMS Symbol fonts on the system, using the
                'amssymb' package.^^J}%
       \usepackage{amssymb}%

      }{}
  \fi

  \IfFileExists{amsbsy.sty}
    {\typeout{^^JFound the 'amsbsy' package on the system, using it.^^J}%
     \usepackage{amsbsy}}
    {}

\newsavebox{\astrutbox}
\sbox{\astrutbox}{\rule[-5pt]{0pt}{20pt}}

\newcommand{\la}{\raisebox{-0.5ex}{$\,\stackrel{<}{\scriptstyle\sim}\,$}}
\newcommand{\ga}{\raisebox{-0.5ex}{$\,\stackrel{>}{\scriptstyle\sim}\,$}}

\input psfig.tex

\title[Galaxy evolution in clusters from $z=1$ to $z=0$]
      {Galaxy evolution in clusters from $z=1$ to $z=0$}

\author[S. Andreon {\it et al.\/}]%
{S. Andreon$^1$, J. Willis$^2$ \thanks{Present Address: Dep. of Physics and Astronomy, 
Univ. of Victoria, Victoria, Canada\hfill}, 
H. Quintana$^2$, \break
I. Valtchanov$^3$, M. Pierre$^4$ \and F. Pacaud \break}

\affiliation{$^1$INAF--Osservatorio Astronomico di Brera, Milano, Italy \\
[\affilskip]
$^2$Dep. de Astronom\'\i a y Astrof\'\i sica, Pontificia Universidad Cat\'olica de Chile, Santiago, Chile\\[\affilskip]
$^3$ Imperial College, London, UK\\[\affilskip]
$^4$CEA/DSM/DAPNIA, Service d'Astrophysique, Gif-sur-Yvette, France\\}

\pubyear{2004}
\volume{195}
\pagerange{1--8}
\date{?? and in revised form ??}
\jname{Outskirts of Galaxy Clusters: intense life in the suburbs}
\editors{A. Diaferio, ed.}
\begin{document}

\maketitle

\begin{abstract}
The XMM--LSS project is detecting
distant clusters of low mass, quite comparable in mass to the ones 
in the local universe. This allows a direct comparison of galaxy
properties at different redshifts in ``similar'' clusters.
We present here first results on the evolution
of the reddest galaxies in $~25$ clusters/groups at $0.3\la z \la 1.0$
and for the whole galaxy population in the same clusters.
The emerging picture from the current study is that the
counterparts of present day clusters tend to show 
two or more distinct populations :
a relatively old ($z_f>2-5$) population evolving passively 
together with a younger population, ostensibly responsible for the apparent
brightening of the characteristic magnitudes, $m^*$.
\end{abstract}

\firstsection 
\section{Introduction}

The nature and the time scale of the processes that shape
galaxies properties in clusters and groups is still unclear, in
spite of significant progress in the past years.

The window opened by the redshift dependence of the galaxies
properties has been used to give constraints on the time scales of the
processes (e.g. Butcher \& Oemler 1984; Dressler et al. 1997;
Stanford, Eisenhardt \& Dickinson, 1998; Treu et al. 2003). However,
many of the clusters compared at different redshifts have  different
masses (or X-ray luminosities), in such a way that in some occasions
``we are comparing unripe apples to ripe oranges in understanding how
fruit ripens" (Andreon \& Ettori 1999). Furthermore, many of
intermediate redshift clusters are optically selected, with a risk of 
biasing the optical properties of galaxies 
(see, for example, Andreon, Lobo \& Iovino, 2004). Finally,
rich clusters are rare environments where evolution is thought to be
accelerated (Kauffman 1996).

There is therefore a compelling need to study galaxy properties of low mass
(much below Coma) clusters at intermediate redshift ($0.3\la z \la 1$). They
have masses similar to common clusters in the local universe.
Selecting them independently from the optical will limit the risk of bias.
This is one of the aims of the XMM--Large Scale Structure
collaboration\footnote[3]{see  http://vela.astro.ulg.ac.be/themes/spatial/xmm/LSS/
\hfill} (Pierre et al. 2004, Andreon, Pierre et al. 2003) that takes advantage
from the multiwavelengh observations in progress on XMM, CTIO,
CFHT Legacy Survey, UKIDSS, Spitzer, Galex and other facilities.  Figure 1 shows individual X--ray
luminosities of more than 1300 clusters of galaxies (black points) drawn from
literature (and listed in the BAX database\footnote[4]{http://bax.ast.obs-mip.fr/
\hfill}). Their median X-ray luminosity (spline) increases with redshift, as shown
in Andreon \& Ettori (1999) for the Butcher \& Oemler (1984) sample. There is,
instead, a good match in X--ray luminosity between XMM--LSS  clusters (the region
at $z\ga0.3$) and the clusters in the local Universe (considered, say, in the
morphology--density paper by Dressler 1980 or  in the Butcher--Oemler paper by
Butcher \& Oemler 1984).

In a single $\sim5$ deg$^2$ area XMM--LSS has currently spectroscopically
observed some 70 clusters at $z\ga0.3$, the most distant being at $z=1.05$ for
the time  being\footnote[5]{A true color image with X-ray contours of the
latter can be inspected at the URL
http://www.brera.mi.astro.it/$\sim$andreon/g04\_06trico.png}. A high redshift
($z>1$) cluster sample is also being assembled, characterized by extended
X--ray emission associated with faint or absent optical galaxies and
significant galaxy overdensity in deep NIR images. The sample it is currently 
composed by at least six $z>1.0$ clusters. 

\begin{figure}
\begin{minipage}[t]{8truecm}
\psfig{figure=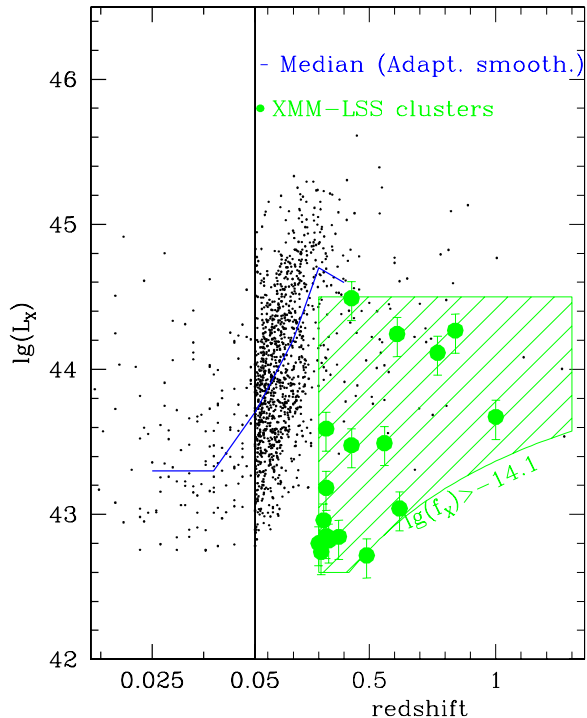,width=8truecm}
\end{minipage}
\quad \vbox{\hsize=5truecm
\caption[]{X-ray luminosity of about 1300 literature clusters (small points),
their median average (spline) and the region accessible to XMM--LSS,
given the 0.8 10$^{-14}$ erg s$^{-1}$ cm$^{-2}$ flux limit in the [0.5-2] keV
band. Green points are clusters confirmed during the 2002 spectroscopic
campaign. }
}
\end{figure}


Here we report results for $\sim20$ clusters at $z\la1$. Most
of them are X--ray selected, with intermediate--low mass (as measured by
their X--ray luminosity)  and low richness $R\la0$, as measured on the
Abell (1958) scale. In addition, we have included a
few, more massive clusters drawn from literature, and a few clusters 
below the X-ray detection limit but
color detected with our own version of the Gladders \& Yee (2000)
method (see Andreon 2003 for an early application to SDSS
data) in the same XMM-LSS sky area. This analysis allow us to check
the impact of the X--ray selection.
More details can be found in Andreon et al. (2004).

\begin{figure}
\begin{minipage}[t]{8truecm}
\psfig{figure=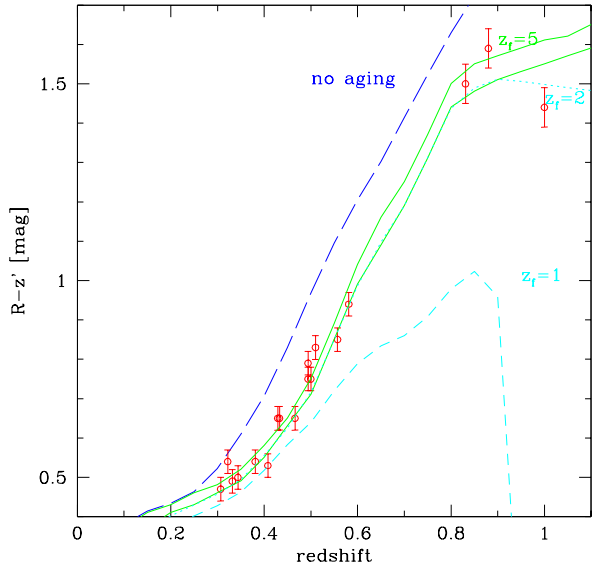,width=8truecm}
\end{minipage}
\quad \vbox{\hsize=5truecm
\caption[]{Observed $R-z'$ colour of the red envelope of the red
sequence observed in each cluster as a function of redshift. Three
galaxy evolution models are considered: a non--evolving early--type
galaxy of present--day age at all redshifts, and two evolving
early--type galaxy models, each characterised by a different formation
redshift and mass.}
}

\end{figure}

\section{The colour--magnitude relation}
Figure 2 shows the colour evolution of the red envelope of the red
sequence, defined as the median colour of the three brightest galaxies
on the red sequence. Several model predictions are
indicated\footnote[5]{Throughout this paper we assume $\Omega_M=0.3$,
$\Omega_\Lambda=0.7$ and $H_0=70$ kms$^{-1}$ Mpc$^{-1}$.}. The top
(long dashed) curve neglects ageing of the stellar population. The
additional models are more physically motivated. Passive stellar
ageing and chemical evolution are described employing the model of
Kodama \& Arimoto (1997) and assume a formation redshift, $z_f$, and a
total stellar mass. The two continuous green curves indicate $z_f=5$
and a total stellar mass of $\sim 1.7 \ 10^{11} M_\odot$ and $\sim 6.4
\ 10^{10} M_\odot$. The expectation for a mass of $\sim 6.4 \ 10^{10}
M_\odot$ and two lower formation redshifts ($z_f=2$ and $z_f=1$) are
plotted as dotted and short dashed curves, respectively. The colour of
the envelope of the red sequence is reproduced well by models where
the stars in the oldest galaxies in the clusters form at $2\la z \la5$
in good agreement with the findings of Stanford et al. (1998) and
Kodama et al. (1998) based on a set of richer clusters located within
a comparable redshift range, and with Andreon et al. (2003) for a
sample of low redshift clusters of low richness. At the difference of
previous works, our observations are quite uniform (virtually all 
photometric data were obtained at the same telescope+instrument in one
single observing run). Furthermore, the cluster sample is mainly selected
independently of the optical properties and avoids a
potentially circular analysis between optically selected clusters and
the properties of the red sequence. Finally, many of the clusters have
low mass, right the mass where current predictions regarding
the assembly of bright, red galaxies (e.g. Kauffmann 1996 and Eggen,
Lynden-Bell \& Sandage 1962) display the greatest divergence.

Therefore, in these clusters, including the low mass ones, the color
magnitude relation zero--point evolves passively, 
i.e. these galaxies must have formed the bulk of their
stellar mass at $z>2-5$. 

\begin{figure}
\psfig{figure=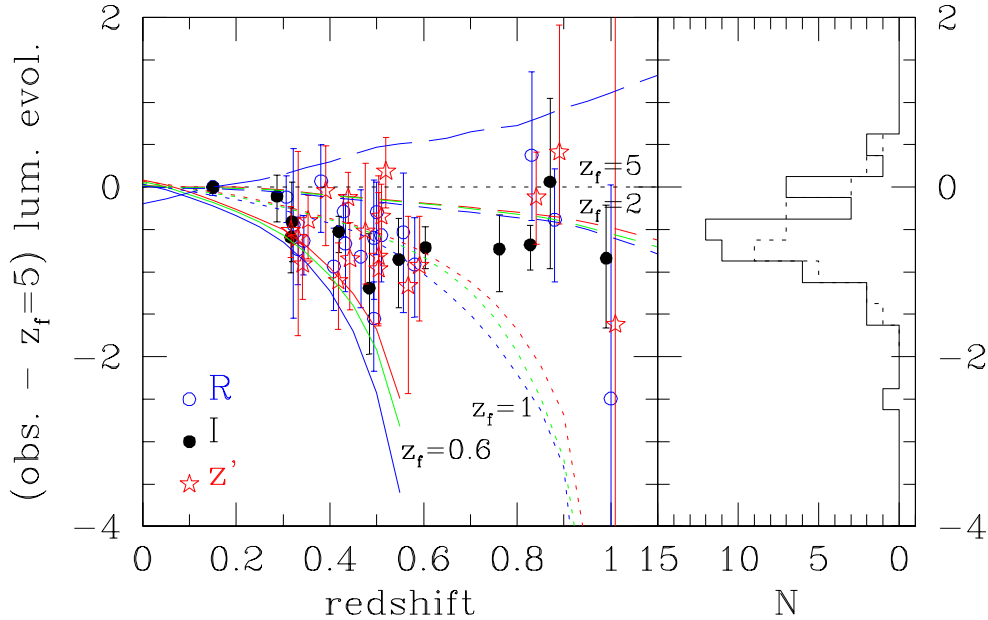,width=13.3truecm}
\caption[h]{{\it Left panel:} Characteristic LF magnitude, $m^*$,
evolution as a function of redshift having removed the constribution
from passive ($z_f=5$) stellar evolution. The labeled curves are the
predictions for different formation redshifts. Points and curves of
the same colour refer to the same filter, as indicated within the
Figure. The long--dashed curve is the expectation neglecting $R$--band
stellar evolution. Points are slightly offset in redshift in order to limit
crawding. {\it Right panel:} frequency distribution of the
points in the left panel (solid histogram) and of the corresponding
values derived without any colour selection and excluding problematic
clusters.}
\end{figure}

\section{The luminosity function}

But what about the whole galaxy population, including
blue galaxies? To this end, we measured the luminosity function
in the standard way, and we fit a Schechter (1978) function of
fixed slope to the data, hence determining $m^*$ and their error.
Figure 3 displays the redshift dependence of 47 $m^*$ values at
$z\ga0.3$ (two $z=0.15$ calibrating points are also plotted, each
one being the average of 21 X--ray selected clusters). To
highlight the possible effects of active luminosity evolution upon
$m^*$, the distance modulus and passive luminosity evolution
terms were removed employing a $m^*$--redshift evolution term
computed for a passively evolving stellar population formed at $z_f=5$
using the model of Kodama \& Arimoto (1997) previously employed to
compute the colour of the red sequence. The model predictions are
normalized to the observed $m^*$ at $z=0.15$. 
The $m^*$ data points are systematically brighter than a model
based upon an old, passively evolving stellar population (horizontal
line). A non--evolving model is also strongly ruled out (rising line).
The clusters
presented in Figure 3 require a secondary star formation episode at
$z_f<5$ in order to
generate an $m^*$ values brighter than the passive evolution model.
The same conclusion can be drawn by considering the right panel of
Figure 3 (solid line) which displays the histogramme of $m^*$
values marginalized over redshift. The resulting $m^*$ distribution is
approximately 1 magnitude wide and is offset from zero toward bright
magnitudes

The additional curves in Figure 3 indicate the expected $m^*$
evolution for stellar populations formed at successively lower
redshifts. In order to account for the bright $m^*$ values observed at
redshifts $z \sim 0.3$ a formation redshift as low as $z_f=0.6$ would
be required, although, by adopting such a low formation redshift, the
predicted $m^*$ value at a redshift $z=0.15$ would be 0.2 magnitudes
brighter than that reported by GMA99. The LF data points are not well
described by any formation model based upon a single episode of star
formation and a more complex scenario should be considered.
The last (in cosmic time) star formation event should
brighten average $m^*$ values by up to 1 mag (right panel of Figure
3). Such secondary star formation activity may be related to the
Bucther--Oemler effect (Bucher \& Oemler 1984), although the evidence
for the latter is not compelling (Andreon \& Ettori 1999; Andreon,
Lobo, Iovino 2004). A Bucther--Oemler analysis of the present sample
of clusters is presently in progress.

\section{Conclusions}

We are at last observing the high redshfit counterparts of the well studied low
redshift clusters. The emerging picture from the current study is that even low
mass clusters at intermediate redshift are composed by two or more distinct
galaxy populations : a relatively old population formed at $2\la z_f \la 5$ and
evolving passively  together with a younger population, ostensibly responsible
for the apparent brightening of the characteristic LF magnitudes.  The
determination of the nature of this secondary activity (e.g. the time scale and
the relationship with the cluster properties) is within the reach of the
XMM--LSS project, since the $z<1.3$ redshift regime is to be ultimately sampled
by several hundreds of X--ray or color selected clusters, with supporting
multi-color and spectroscopic observations.

\begin{acknowledgments}
SA thanks the whole XMM-LSS collaboration for the numerous
discussions. SA wormly thanks A. Dressler for useful suggestions.
This work has received support from MURST-COFIN n. 2003020150-005
\end{acknowledgments}


\begin{thebibliography}{}


\bibitem[Abell(1958)]{1958ApJS....3..211A} 
Abell, G.~O.\ 1958, ApJS, 3, 211 

\bibitem[Andreon(2003)]{2003A&A...409...37A} 
Andreon, S.\ 2003, A\&A, 409, 37 

\bibitem[Andreon \& Ettori(1999)]{1999ApJ...516..647A} 
Andreon, S.~\& Ettori, S.\ 1999, ApJ, 516, 647 

\bibitem[]{} 
Andreon, S., Lobo, C., Iovino A., \ 2004, MNRAS, 349, 889

\bibitem[Andreon, Pierre, \& the XMM-LSS 
collaboration(2003)]{2003MSAIS...3..188A} Andreon, S., Pierre, M., \& the 
XMM-LSS collaboration 2003, Societa Astronomica Italiana Memorie 
Supplement, 3, 188

\bibitem[]{}
Andreon, S., Willis, J., Quintana, H., Valtchanov, I., Pierre, M., 2004,
MNRAS, submitted

\bibitem[Butcher \& Oemler(1984)]{1984ApJ...285..426B} 
Butcher, H.~\& Oemler, A.\ 1984, ApJ, 285, 426 

\bibitem[Dressler et al.(1997)]{1997ApJ...490..577D} 
Dressler, A., et al.\ 1997, ApJ, 490, 577 


\bibitem[Gladders \& Yee(2000)]{2000AJ....120.2148G} 
Gladders, M.~D.~\& Yee, H.~K.~C.\ 2000, AJ, 120, 2148 

\bibitem[Eggen, Lynden-Bell, \& Sandage(1962)]{1962ApJ...136..748E} 
Eggen, O.~J., Lynden-Bell, D., \& Sandage, A.~R.\ 1962, ApJ, 136, 748 


\bibitem[Kauffmann(1996)]{1996MNRAS.281..487K} 
Kauffmann, G.\ 1996, MNRAS, 281, 487 


\bibitem[Kodama \& Arimoto(1997)]{1997A&A...320...41K} 
Kodama, T.~\& Arimoto, N.\ 1997, A\&A, 320, 41

\bibitem[]{}
Pierre et al. 2004, PASP, submitted (astro-ph/0305191)

\bibitem[Schechter(1976)]{1976ApJ...203..297S} 
Schechter, P.\ 1976, ApJ, 203, 297 

\bibitem[Stanford, Eisenhardt, \& Dickinson(1998)]{1998ApJ...492..461S} 
Stanford, S.~A., Eisenhardt, P.~R., \& Dickinson, M.\ 1998, ApJ, 492, 461 

\bibitem[Treu et al.(2003)]{2003ApJ...591...53T} Treu, T., Ellis, R.~S., 
Kneib, J., Dressler, A., Smail, I., Czoske, O., Oemler, A., \& Natarajan, 
P.\ 2003, ApJ, 591, 53 

\end{thebibliography}
\end{document}